\documentclass[english,twocolumn]{revtex4-1}
\usepackage[T1]{fontenc}
\usepackage[latin9]{inputenc}
\usepackage{amstext,amssymb,amsmath}
\usepackage{graphicx}
\usepackage{esint}
\usepackage{babel}
\usepackage{tikz}

\begin{document}

\title{Mode hopping in oscillating systems with stochastic delays}

\author{Vladimir~Klinshov, Dmitry Shchapin}

\affiliation{Institute of Applied Physics of the Russian Academy of Sciences,
46 Ul'yanov Street, 603950, Nizhny Novgorod, Russia}

\author{Otti D'Huys}

\affiliation{Department of Mathematics, Aston University, B4 7ET Birmingham,
United Kingdom}

\begin{abstract}
We study a noisy oscillator with pulse delayed feedback, theoretically and in an electronic experimental implementation. Without noise, this system has multiple stable periodic regimes. We consider two types of noise: i) phase noise acting on the oscillator state variable and ii)  stochastic fluctuations of the coupling delay. For both types of stochastic perturbations the system hops between the deterministic regimes, but it shows dramatically different scaling properties for different types of noise.  The robustness to conventional phase noise \textit{increases} with coupling strength. However for stochastic variations in the coupling delay, the lifetimes \textit{decrease} exponentially with the coupling strength. We provide an analytic explanation for these scaling properties in a linearised model. Our findings thus indicate that the robustness of a system to stochastic perturbations strongly depends on the nature of these perturbations.
\end{abstract}

\date{\today}

\maketitle

Many networks of various nature exhibit temporal delays accounting for the traveling time of a propagating signal. Coupling delays arise in laser physics \cite{kozyreff2000global,yanchuk2004instabilities,fischerRevModPhys}, neuroscience \cite{ko2007effects,vicente2008dynamical}, gene regulatory networks \cite{mackey1977oscillation,danino2010synchronized}, traffic and population dynamics \cite{kuang1993delay,nagatani1998delay}, communication networks \cite{leon2003communication}, etc. A typical effect of time delays is multistability: different types of dynamics are possible for the same parameter values \cite{foss2000multistability,yanchuk2009delay,klinshov2015multistable,klinshov2015emergence}. Hence, stochastic perturbations, which are present in any real-life system, can cause hopping between coexisting stable states \cite{kramers1940brownian,ikeda1989maxwell,wiesenfeld1989attractor,arecchi1990experimental,
kraut2002multistability}.

In this Letter, we investigate the hopping dynamics in the most basic time-delayed network: a single node with delayed feedback. We utilize a model of a phase oscillator with pulse coupling which is widely used for biological oscillators \cite{winfree2001geometry,peskin1975mathematical,lewis1992phase}, wireless networks \cite{pagliari2010bio}, chemical and electronic oscillators \cite{lopera2006ghost,rosin2013control,safonov2017dynamical} and optical systems \cite{colet1994digital}. We consider two different types of stochastic perturbations. First, we consider ``phase noise'', an additive stochastic term in the equation for the oscillator phase. This is the canonical implementation of stochastic effects \cite{kramers1940brownian,lindner2004effects,ren2010noise,zakharova2010stochastic,franovic2015activation}. Second, we investigate the influence of noise not on the node, but on the link by letting the coupling delay fluctuate. 

Stochastic delays are rarely taken into account due to the mathematical complexity of their implementation  \cite{verriest2009deterministic,sadeghpour2018can}. Nevertheless, they arise naturally in various systems such as gene regulatory networks \cite{josic2011stochastic}, networked control and communication systems \cite{nilsson1998stochastic,lin2009observer,qin2017stability} and networks of electronic gates \cite{dhuys2016}.  Fluctuations in the coupling delay may significantly influence network dynamics: for example they may deteriorate the performance and stability of communication networks \cite{krtolica1994stability} or increase signaling speed in gene regulatory networks \cite{josic2011stochastic}. 

While most research so far focuses on the linear stability of discrete-time \cite{gomez2013stability,qin2017stability} or continuous-time systems \cite{verriest2009stability,gomez2016stability}, our major interest is to study the switching between different attractors of the deterministic system induced by the fluctuations. In the following, we provide a numerical study of a pulse-coupled oscillator together with an electronic experiment. Moreover, we develop an analytic theory describing the noise-induced switching between different states. Our main result is that the robustness of the system towards both types of stochastic perturbations -- in the oscillator or in the coupling  delay --  follows opposite scaling laws: an increased coupling strength makes the system less susceptible to phase noise, but at the same time it becomes more sensitive to the fluctuations of the delay. 

We consider the following model of a single phase oscillator with pulse delayed feedback
\begin{equation}\label{eq1}
\frac{d\phi}{dt}=1+\epsilon Z(\phi)\sum_{t_s} \delta(t-t_s-\tau)\,.
\end{equation}
The motion along the limit cycle is modeled as $d\phi/dt=1$. When the phase reaches unity, it resets to zero and the oscillator emits a spike. The moments of spike emission are denoted as $t_s$. The spike is received after a delay $\tau$, at a reception phase $\psi=\phi(t_s+\tau)$, and causes an instantaneous shift $\Delta\phi=\epsilon Z(\psi)$, where $\epsilon$ is the coupling strength and $Z(\phi)$ is the phase response curve (PRC, \cite{canavier2010pulse}). 

The deterministic dynamics of this system \eqref{eq1} has been studied in detail in Ref. \cite{klinshov2015emergence}. The basic regime of the system is the so-called regular spiking regime, characterized by a constant inter-spike interval (ISI) between consecutive spikes. This regime is characterized by a capacity $C$, the number of full inter-spike intervals within a delay interval. The oscillator receives exactly one pulse in each ISI. The deterministic reception phase $\psi_C$, the deterministic period $T_C$ and the delay $\tau$ are related by
\begin{equation}\label{eq:T_C}
\psi_C=\tau-CT_C,\quad T_C  =  1-\epsilon Z(\psi_C).
\end{equation}
For large enough delays several regular spiking solutions with different capacities coexist, each characterized by different reception phase $\psi_C$ and period $T_C$. The stability condition of each solution is given by $-1<\epsilon Z^\prime(\psi_C)<1/C$. Their number grows with the delay and the coupling strength,  $N\sim \epsilon\tau$, and their difference decreases as $T_{C-1}-T_C \sim \frac{1}{\tau}$.

We consider two different types of stochastic perturbations to system \eqref{eq1}:

i) The stochastic perturbation is applied to the oscillator as an additive noise term: 
\begin{equation}\label{eq1a}
\frac{d\phi}{dt}=1+\epsilon Z(\phi)\sum_{t_s} \delta(t-t_s-\tau)+\sigma_p\xi(t)\,.
\end{equation}
Here, $\xi(t)$ is standard white Gaussian noise, and $\sigma_p$ is the noise strength. We refer to this scenario as ``phase noise''. We integrate this system (Eq. \eqref{eq1a}) using an Euler-Mayurama scheme \cite{baker2005exponential} with a step size of $dt=10^{-3}$.

ii) The perturbation is applied to the coupling delay. Stochastic delays are implemented by adding random uncorrelated variations $\sigma_d\xi_s$ to the delay of a pulse emitted at $t_s$:
\begin{equation}\label{eq1b}
\frac{d\phi}{dt}=1+\epsilon Z(\phi)\sum_{t_s} \delta(t-t_s-\tau+\sigma_d\xi_s)\,.
\end{equation}
where $\xi_s$ is a discrete standard normally distributed noise term, and $\sigma_d$ measures the fluctuations intensity. We always chose the noise strength small enough and truncate the distribution, so that the delay remains positive and the pulse order preserves. Taking advantage of the discrete nature of the coupling, the system Eq. \eqref{eq1b} is integrated using an event-based method.

%
%

\begin{figure}[!h]
\centering
\includegraphics[width=0.45\textwidth]{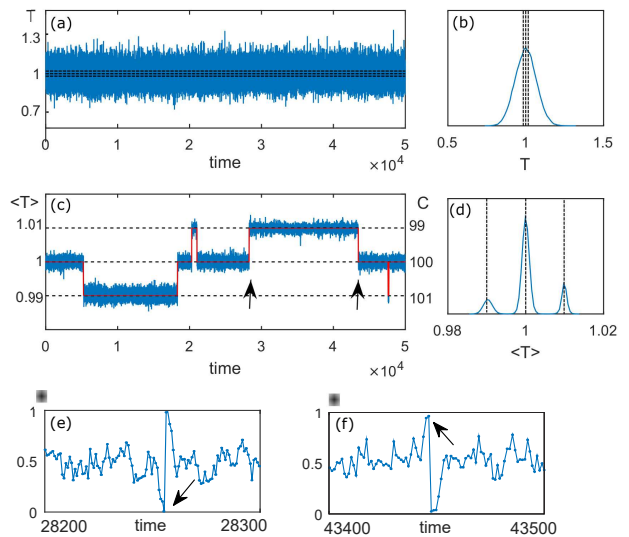}
\caption{Noise-induced switching. (a) Temporal dynamics and (b) the distributions of the inter-spike intervals. Black dashed lines correspond to the periods of the deterministic solutions. (c) Temporal dynamics of the ISIs after applying a moving average filter with a width $\tau$ (blue) and the capacity (red). (d) Distribution of the filtered ISIs. (e) The dynamics of the input phase $\psi$ when the system switches to the state with the lower capacity. (f) The same for the switching to the higher capacity. The switching moments are indicated by arrows. The system parameters: phase noise, $\tau=100.5$, $\epsilon=0.1$, $\sigma_p=0.06$. }
\label{fig:switching}
\end{figure}

In the following we compare those two scenarios. We select large enough delay so that the system is multistable, and analyse the noise-induced switching between different regimes of regular spiking (Fig. \ref{fig:switching}). As a PRC, we choose
\begin{equation}\label{eq2}
Z(\phi)=\frac{1}{2\pi}\sin(2\pi\phi)\,,
\end{equation} 
which is characteristic for oscillators close to Hopf bifurcation \cite{brown2004phase}. We choose the coupling delay $\tau=C_0+\frac{1}{2}$, so that there is always a central solution with capacity $C_0$, reception phase $\psi_{C_0}=\frac{1}{2}$ and period $T_{C_0}=1$. This choice of delay does not affect the general switching properties, but facilitates the comparison of the switching statistics for varying delay and coupling strength. For weak enough coupling, this central solution with $\psi_{C_0}=\frac{1}{2}$ is the most stable one, and it is surrounded by two unstable solutions with $\psi_C\approx 0$ and $\psi_C\approx 1$. 

The natural, and in experiments the only observables of our system, are the spike times and the inter-spike intervals $T$. However, for both types of stochastic perturbations neither the time series of the consecutive ISIs (Fig. \ref{fig:switching}a), nor the distribution of ISIs (Fig. \ref{fig:switching}b) reveal any signs of switching. Clearly, for large delays the difference between the periods of neighboring regimes $T_C -T_{C\pm 1}$ is much smaller than the typical ISI fluctuations. Nevertheless, after applying a moving average filter with a width $\tau$ one sees clearly pronounced mode hopping (Fig. \ref{fig:switching}c), while the distribution of filtered ISIs $\langle T \rangle$ shows multiple peaks corresponding to different regular regimes (Fig.\ref{fig:switching}d). 

In the numerical simulations, the switching events can also be inferred from the dynamics of the reception phases $\psi$.
If the phase decreases and passes the unstable state at $\psi\approx 0$, a switch to a solution with lower capacity is observed (Fig. \ref{fig:switching}e). If the phase $\psi$ increases and passes through the unstable state $\psi\approx 1$, this coincides with a switch to a solution with a higher capacity (Fig. \ref{fig:switching}f). Thus, the variable $\psi +C = \phi(t_s+\tau)-\phi(t_s)$, which corresponds to the delay phase difference at the reception of the spike, provides an indicator for the switching dynamics. Because of our choice of a sinusoidal PRC, with the unstable states located around 0 and 1, the capacity $C$ itself is a straightforward indicator of the regime in which the system resides (cf. Fig. \ref{fig:switching}c).

\begin{figure}[!h]
\centering
\includegraphics[width=0.45\textwidth]{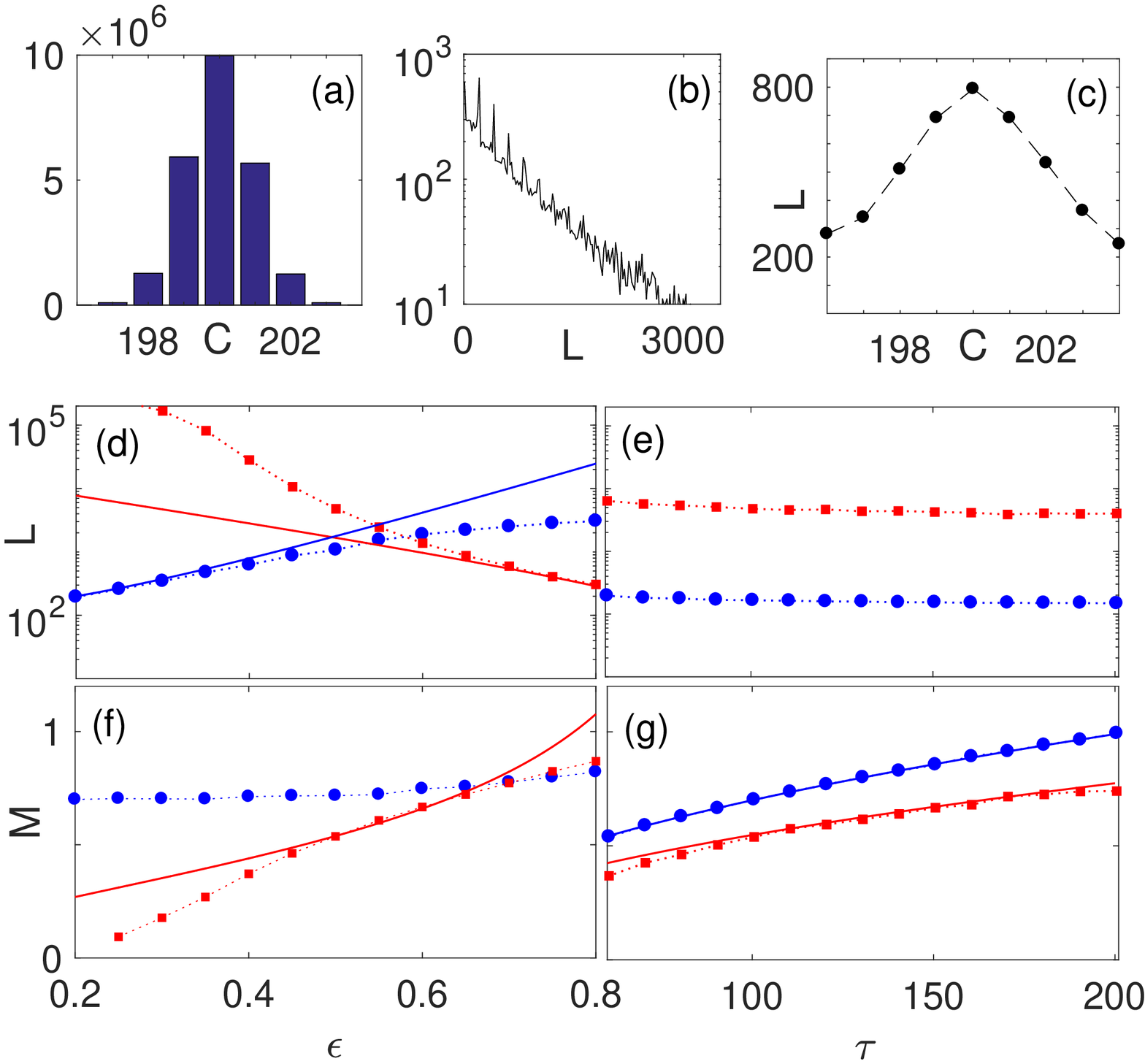}
\caption{An exemplary capacity distribution (a), distribution of the lifetime of the central solution (e), and the average lifetimes of the solutions with different capacity (c) due to mode hopping. Parameters for (a-c): stochastic delays, $\epsilon=0.65$, $\tau=200.5$ and $\sigma_d=0.17$. 
\newline Panels (d-f) show the characteristics of switching for phase noise (blue circles) and stochastic delays (red squares). The average lifetime $L$ of the central solution (d,e) and the width $M$ of the capacity distribution (f,g) versus the coupling strength $\epsilon$ (d,f) and the delay $\tau$ (e,g) Parameters are $\sigma_p=0.07$, $\sigma_d=0.17$ and $\tau=100.5$ (d,f) and $\epsilon=0.15$ (phase noise) and $\epsilon=0.5$ (stochastic delays) (e,g). The full lines correspond to the scaling laws (\ref{eq:LM-phase}) (blue, for phase noise) and  (\ref{eq:L-delay}) and (\ref{eq:M-delay}) (red, for stochastic delays). }\label{fig:main}
\end{figure}

We characterize the mode hopping statistics by two measures. First, we are interested in the the typical number of states with different capacity that are visited; an exemplary distribution of capacities is shown in Fig. \ref{fig:main}a. This number is related to the standard deviation $M$ of the distribution of the capacities, $M=\sqrt{\langle C^2\rangle-\langle C\rangle^2}$. Second, the lifetimes of the stable states provide an indication of the robustness of the system to noise and the memory capacity. We find that the lifetimes are distributed exponentially for both types of noise, with additional peaks at multiples of the delay time, as can be expected \cite{kramers1940brownian,masoller2002,dhuys2014stochastic}.  An exemplary distribution is shown in Fig. \ref{fig:main}b. The lifetime are typically maximal for the central state with capacity $C_0$, and lower for the other states, the average lifetimes for states with different capacity is shown in Fig. \ref{fig:main}c. We consider the average lifetime $L$ of the central state as the temporal characteristic of the switching.

Figure \ref{fig:main} (d-g) compares the numerical results for phase noise (Eq. \eqref{eq1a}) and stochastic delays (Eq. \eqref{eq1b}) applied to the oscillator. For phase noise, the width $M$ of the capacity distribution increases with the delay time $\tau$ and only weakly depends on the coupling strength $\epsilon$. The lifetime $L$ weakly depends on the delay and increases quickly with the coupling strength. It is instructive to compare these numerical results with the theory for continuous coupling \cite{dhuys2014stochastic}, which predicts the following scaling for large delays:
\begin{equation}
L \sim \frac{1}{\epsilon} \mbox{exp}\left(\frac{\epsilon}{2\pi^2\sigma_p^2}\right),\quad M\sim \sigma_p \sqrt{\tau}.
\label{eq:LM-phase}
\end{equation}

This predicted scaling (indicated by the full blue lines in Fig. \ref{fig:main} (d-g)) is in good agreement with our results for weak coupling when pulsatile coupling can be approximated by continuous coupling \cite{goel2002synchrony}. As the coupling increases, the approximation is no longer valid, and the agreement with the scaling laws \eqref{eq:LM-phase} deteriorates.


For stochastic delays, similarly to phase noise, the distribution width $M$ scales as $\sqrt{\tau}$, while the lifetime $L$ does not change much with $\tau$. However, the difference in the role of the coupling strength $\epsilon$ is striking. In contrast to phase noise, under influence of stochastic delays the distribution width $M$ grows with $\epsilon$, while the lifetime $L$ decreases almost exponentially. These results suggest that strengthening the coupling unexpectedly makes the system more susceptible to stochastic fluctuations of the delay.

In order to explain these numerical results we develop a theory, based on the PRC (Eq. \eqref{eq2}) linearized around the central phase $
z(\phi)=\frac{1}{2}-\phi$. A linear approximation of the PRC will not provide a quantitative explanation about the mode hopping characteristics, as these depend on system behavior far from the deterministic solutions. Nevertheless, the analysis with a linear PRC allows insight into the scaling laws governing the mode hopping due to stochastic delays. 
For large delays $\tau\gg \epsilon^{-1}$, the stable regimes close to the central solution are approximated as $\psi_{C}\approx\frac{1}{2}-\frac{C-C_0}{\epsilon C_0}$, $T_{C}\approx 1-\frac{C-C_0}{C_0}$. 

We start by rewriting Eq. \eqref{eq1b} as a stochastic map. Assuming a capacity $C$ at the time $t_s$ of a pulse emission, the next pulse to arrive was emitted at $t_{s-C}.$ The reception phase $\psi_{s+1}$ (the index refers to the spike time following this pulse) is then given by the time elapsed between $t_s$ and the arrival of this pulse:
\begin{equation}\label{eq:psis}
\psi_{s+1}=\tau+\sigma\xi_{s+1}-(t_{s}-t_{s-C})=\tau-\sum_{p=s-C+1}^{s}T_p+\sigma\xi_{s+1},
\end{equation}
where $T_p=t_{p}-t_{p-1}$ are the inter-spike intervals. The phase grows uniformly except at the moment of the pulse reception, and the oscillator receives one pulse per inter-spike interval, thus the inter-spike interval equals $T_{s+1}=1-\epsilon z(\psi_{s+1})$. It is convenient to introduce the deviation of the reception phase from its steady state value $x_s=\psi_s-\psi_C$, then the stochastic map Eq. (\ref{eq:psis}) can be rewritten as
\begin{equation}\label{eq:xs}
x_{s+1}=-\epsilon\sum_{p=s-C+1}^{s}x_p+\sigma_d\xi_s.
\end{equation}
This is an autoregressive process of order $C$. Using the Yule-Walker equations \cite{chatfield}, it is straightforward to show that $\langle x_s \rangle=0$ and to calculate the variance and the autocorrelation coefficients:
\begin{eqnarray}
v&=&\langle x_s^2\rangle=\sigma_d^2\frac{1+(C-1)\epsilon}{1+(C-1)\epsilon-C\epsilon^2},\\
\rho_n&=&\frac{\langle x_sx_{s-n}\rangle}{\langle x_s^2\rangle}=-\frac{\epsilon}{1+(C-1)\epsilon} \mbox{ for }1\leq n\leq C.
\end{eqnarray}

The typical capacity $C$ is close to the delay, as can be deduced from Eq. \eqref{eq:T_C}; hence for large delays the autocorrelation coefficients vanish, while the variance tends to the limit 
$v=\frac{\sigma^2_d}{1-\epsilon}$. Thus, the reception phases $\psi_s$ are, approximately, normally distributed around the deterministic value $\psi_C$ with a variance $v$. A hopping to a different solution corresponds to the reception phase crossing the boundary  $\psi_s=1$ (if the capacity increases) or $\psi_s=0$ (if the capacity decreases). The switching statistics can be derived from solving the first passage time problem, which is not trivial for an autoregressive process \cite{di2008first}. However, due to the low correlation for large delays it is possible to estimate the switching rate as the probability to find the value of $\psi_s$ beyond the boundary:
\begin{eqnarray}
r_+(C)&\approx&P(\psi_s>1)=1-\Phi\left(\frac{1-\psi_C}{\sqrt{v}}\right),\\
r_-(C)&\approx&P(\psi_s<0)=\Phi\left(\frac{-\psi_C}{\sqrt{v}}\right).
\end{eqnarray}
where $\Phi(x)=\frac{1}{2}(1+\mbox{erf}(x/\sqrt{2}))$. For weak noise these probabilities are small, and using an approximation for the tails of the error function, we obtain 
\begin{equation}
r_{\pm}\approx\frac{\sqrt{v}}{2\sqrt{\pi}\Delta_\pm}\exp\left(-\frac{\Delta_\pm^2}{2v}\right),
\end{equation}
where $\Delta_+=1-\psi_C$ and $\Delta_-=\psi_C$ are the distances from $\psi_C$ to $1$ or $0$, respectively. The average lifetime equals $L(C)=\left((r_+(C)+r_-(C)\right)^{-1}$, and for the solutions not far from the central one ($\psi_C-\frac{1}{2}\ll 1$) it can be approximated by

\begin{equation}
L(C)\approx\frac{1}{2}\sqrt{\frac{\pi}{v}}\exp\left(\frac{1}{8v}\right)\left[\cosh\left(\frac{C-C_0}{2\epsilon v\tau}\right)\right]^{-1}.
\end{equation}

Thus, the average lifetime $L$ of the central solution is given by

\begin{eqnarray}
L&\sim&\frac{\sqrt{(1-\epsilon)}}{\sigma_d}\exp\left(\frac{1-\epsilon}{8\sigma_d^2}\right),
\label{eq:L-delay}
\end{eqnarray}
which provides a qualitative explanation for the almost exponential decrease of $L$ with $\epsilon$ shown in Fig. \ref{fig:main}d.

From the switching rates one can calculate the distribution of the capacities:  assuming detailed balance $p(C)r_+(C)=p(C+1)r_-(C+1)$ we find
\begin{equation}
p(C)\approx p(C_0)\exp\left(-\frac{(C-C_0)^2}{2\epsilon v \tau}\right).
\end{equation}
This is a discrete normal distribution whose standard deviation can be estimated as \cite{szablowski2001discrete}
\begin{eqnarray}
M\sim\sqrt{\epsilon v \tau}=\sigma_d\sqrt{\frac{\epsilon\tau}{1-\epsilon}}.
\label{eq:M-delay}
\end{eqnarray}
The numerical results shown in Fig. \ref{fig:main}(f,g) indeed show a square root scaling of $M$ with $\tau$. While Eq. (\ref{eq:M-delay}) explains the increase of $M$ with $\epsilon$, the correspondence is only qualitative.

Comparing the theoretical approximations for phase noise (Eq. \eqref{eq:LM-phase}) and stochastic delays (Eqs. \eqref{eq:L-delay} and \eqref{eq:M-delay}), we find that the scaling of the switching statistics with the coupling delay and the noise strengths are similar for both types of noise. However, the scaling with the coupling strength is completely different. While the lifetime exponentially increases with $\epsilon$ for phase noise, it exponentially decreases  for stochastic delays. While for phase noise the distribution width $M$ does not depend on $\epsilon$, it increases with $\epsilon$ for stochastic delays. Intuitively the opposite role of the coupling strength for different types of noise can be understood as follows. The fluctuations in the phase due to phase noise only depend on the noise strength $\sigma_p$. When increasing the coupling strength $\epsilon$, these fluctuations are more effectively suppressed since the Lyapunov exponents of the stable periodic states scale with the coupling strength. In contrast, for the stochastic delays the perturbation enters through the coupling, and its effect increases with the coupling strength. 

\begin{figure}[!t]
\centering
\includegraphics[width=0.45\textwidth]{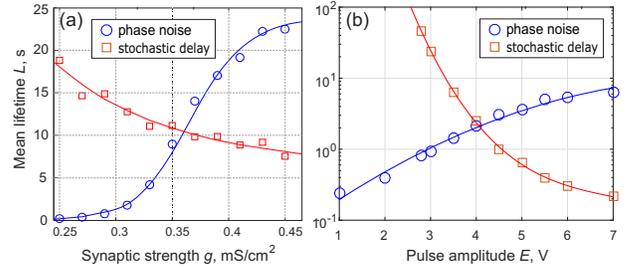}
\caption{Average lifetime of the most stable solution versus the coupling strength for the phase noise (blue circles) and stochastic delay (red boxes) in  (a) a numerically simulated neural oscillator and (b) an experimentally studied electronic circuit. The curves connecting the datapoints are fits. Details are given in the Supplemental materials.}\label{fig:exp}
\end{figure}

For our analysis we have considered delays much larger than the intrinsic period, and a simplified oscillator model; this allows to develop an analytic theory. To show the relevance of our results for more realistic oscillator models and shorter delays, we performed a limited study of a Wang-Buzsaki neuron \cite{wang1996gamma} with an excitatory synapse projecting onto itself (see Supplemental materials for the details). For strong enough synaptic strength $g$, the system is bistable even for synaptic delays shorter than the intrinsic period. The results are presented in Fig. \ref{fig:exp}a and show the same tendency: lifetimes increase with coupling for phase noise but decrease for stochastic delays.

In order to corroborate our theoretical predictions and numerical simulations we carried out an experimental study with an electronic circuit. The experimental setup is described in detail in the Supplemental materials. It was based on the electronic FitzHugh-Nagumo oscillator \cite{shchapin2009dynamics,klinshov2014cross}. When the output voltage exceeds a threshold value, a spike is produced and sent to a delay line. When the spike has passed through the delay line, a short square voltage pulse is applied to the oscillator. To control the coupling strength we varied the pulse amplitude $E$. To implement the stochastic delay, the initial position of each spike in the delay line was randomly shifted. The phase noise was implemented as a white Gaussian noise applied to the oscillator. For either type of the noise, the switching is clearly seen from the multi-peak distribution of the averaged inter-spike intervals (see Fig. S7 in the Supplemental materials). In Fig. \ref{fig:exp}b the average lifetime of the central solution is plotted versus the coupling strength for  both types of noise. In agreement with our theoretical prediction, it grows with the coupling strength for phase noise and decreases for stochastic delay.

The results reported in this Letter show that the effect of stochastic perturbations on oscillatory systems with coupling delay may differ depending on whether they are applied on the oscillator or on the delay line. Our main finding is the unusual scaling properties of the switching dynamics for stochastic delays: the lifetimes \textit{decrease} exponentially with the coupling strength. Our results emphasize the importance of studying stochastic delays which may cause unexpected dynamical effects. In a broader context, a system considered here is the most simple form of a network, with one node and one link. It might also be seen as an analogue of a larger network with ring topology \cite{yanchuk2008destabilization,perlikowski2010periodic,perlikowski2010routes,klinshov2017embedding}.
Our findings suggest that network robustness may strongly depend on whether stochastic perturbations affect the nodes or the links in a network. 

\begin{acknowledgments}
This work is jointly funded by and The Russian Foundation for Basic Research (grant 19-52-10004 for V.K.) and The Royal Society (grant agreement IEC\textbackslash R2\textbackslash 181113 for O.D.). O.D has received funding from the European Union's Horizon 2020 research and innovation programme under the Marie Sklodowska-Curie grant agreement No 713694. V.K. and D.S. acknowledge the support of the Russian Science Foundation (grant 19-12-00338 for the experimental study and grant 19-72-10114 for the numerical simulations).
\end{acknowledgments}

\end{document}